\documentclass[cits]{PoS}

\title{Are the narrow unflavoured mesons, baryons, and dibaryons, a signature of a new physics ?}

\ShortTitle{Are narrow unflavoured mesons a signature of new physics ?}

\author{Boris Tatischeff \\
 Univ. Paris-Sud, IPNO, UMR-8608, and 
 CNRS/IN2P3, Orsay, F-91405, France\\  
E-mail: \email{tati@ipno.in2p3.fr}
}
   
\author{\speaker{Egle Tomasi-Gustafsson}\\
         CEA,IRFU,SPhN, Saclay, 91191 Gif-sur-Yvette Cedex, France\\ 
        E-mail: \email{etomasi@cea.fr}
}

\abstract{New data of narrow low mass unflavoured mesonic structures are presented. A table of these exotic masses is obtained adding previously published data. The mass sequence shows a significant coupling of some of these clusters with stable hadrons: pion, nucleon, and deuteron. Indeed this coupling allows to reproduce rather well the masses of exotic narrow baryons and dibaryons. A discussion is presented  to suggest a possible interpretation of these exotic hadronic structures.}

\FullConference{XXII International Baldin Seminar on High Energy Physics Problems \\
		 September 15-20, 2014  \\
		 JINR, Dubna, Russia
	
}
\begin{document}

\section{Introduction}
Many data show the presence of weakly excited, narrow, exotic low mass structures in unflavoured mesonic spectra. They are called exotic, since there is no room for them inside the quark ${\it q}{\bar q}$ model. Several data have been reported in \cite{bormes}, recalling our previous results, as well as  other results. A partial summary of the results shown in the present paper was recently published \cite{nelmum3}. The first measurements were done using bubble chambers \cite{troyan}. All these data will be discussed below.

With exception of the present Fig. 1 (left), which is reprinted, the other data are read and reproduced in the plotted figures.

Several extracted mesonic structures are small, hardly filling the 5 $\sigma$ criterium. Their justification lies in the existence of similar observations at the same (or nearby) masses. The number of standard deviations (S.D.), in units of $\sigma$, is defined as:\\
\begin{equation}
S.D. =  \sum_{i=1}^n N_i/(\Delta N_i)^2 \hspace*{2.mm}  / \hspace*{2.mm} \sqrt{\sum_{i=1}^n [1/(\Delta N_i)^2]}
\end{equation}
where $\Delta N_i$ corresponds to the total uncertainty, namely the
uncertainty on peak plus background. The uncertainty on
background is taken to be the same as the peak plus background
one. Therefore the obtained S.D. is pessimistic. $N$ stands for the signal, i.e., 
total minus background data.

Beyond the confirmation of the exotic mesonic structures mass spectrum, it remains a challenge for theoreticians to explain their nature. It is presently assumed inside the quark model, that glueballs, hybrids, or ${\it q}{\bar q}$ plus gluon, have masses above M $\approx$ 1.5~GeV. In the same way, the lowest tetraquark  $\pi_{1}$(1400) $1^{-}(1^{-+})$ has mass $M= 1354 \pm 25$~MeV from the Particle Data Group (PDG) \cite{pdg}. So all these mesons have masses above the mass range discussed in the present paper. Possible low mass molecules should be close to two pions (or other two mesons).  
\section{Selection of some data showing low mass narrow mesonic structures}
A selection of data is presented here, either reproducing useful spectra from reactions not discussed up to now, or precise (statistics, resolution, binning) spectra again not always published in peer reviewed journals. 
\subsection{The ${\it p}{\bar p}$ annihilation at rest}

\begin{figure}[h] 
\begin{center}
\scalebox{0.85}[1]{
\hspace*{-0.28cm}
\includegraphics[scale=0.8] {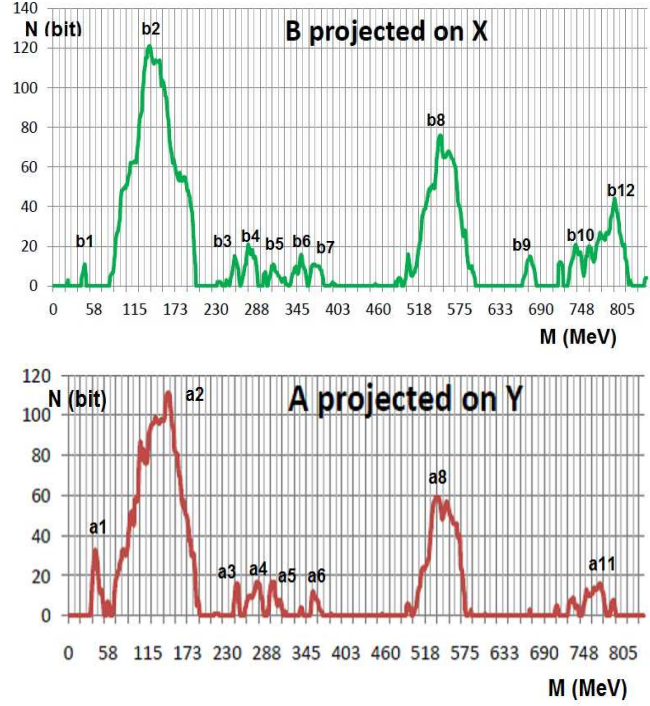}}
\caption{(Color on line). Scanned insert from Fig.~6 of \cite{amsler}  showing the scatter plot of the two photon invariant mass $M(\gamma3, \gamma4)$ (in MeV), versus the two photon invariant mass $M(\gamma1, \gamma2)$ (in MeV).   Reprinted figure with permission from Claude Amsler, Rev. of Modern Physics, Vol. 70, 1293 (1998).}
\label{fig:fig1}
\end{center}
\end{figure}
The  ${\it p}{\bar p}$ annihilation at rest was studied at LEAR (CERN) \cite{amsler} with help of the Crystal Barrel. Since the original data are no more available, Fig. \ref{fig:fig1} showing the scatter plot  of 2$\gamma$ invariant masses  versus   2$\gamma$ invariant masses: M($\gamma_3\gamma_4)$ = f(M($\gamma_1\gamma_2$)) was scanned. The projection are reported in the  Fig. \ref{fig:fig1}, where the intensity of the peaks is proportional to the width of the corresponding bidimensional spot in the original scatterplot. The most intense concentration of data  appears in spots corresponding to $\pi^{0}\pi^{0}$,  $\pi^{0}\eta$, $\pi^{0}\omega$, $\eta\eta$, and $\eta\omega$ coincidences.Bo	 Table 1 shows the masses of the various spots located between the $\pi^{0}\pi^{0}$ and $\pi^{0}\eta$ spots, in the ranges A and B, and projected on both axis X and Y.  A good agreement between these masses is observed, which supports the confidence on their interpretation to be genuine physical peaks and not background.
\begin{table}[h]
\vspace*{-0.2cm}
\caption{Comparison of narrow unflavoured meson masses observed in the  M($\gamma3\gamma4)$ = f(M($\gamma1\gamma2$)) scatter plot from LEAR (CERN) \cite{amsler}. AO means "already observed" from different reactions (see table 2 and Fig. 5). All masses are in [MeV].}
\label{table:table1}
\begin{center}
\begin{tabular}{c c c c c}
\hline
AO&name(1)&mass(1)&name(2)&mass(2)\\
\hline
 &a1&45&a2&45.7\\
81.3&&&h2&85\\
252&b1&252.5&b2&248.5\\
&c1&275&c2&274.7\\    
309.9&d1&310&d2&309.6\\
347&e1&345& &\\
&f1&375&f2&365\\
675.2&g1&670&g2&674\\
\hline
\end{tabular}
\end{center}
\end{table}
\subsection{The ${\it pp} \to {\it pp}\pi^{+}\pi^{-}$ reaction}
\begin{figure}[h] 
\scalebox{0.85}[0.3]{
\hspace*{-0.28cm}
\includegraphics[bb=20 14 517 517,clip,scale=1] {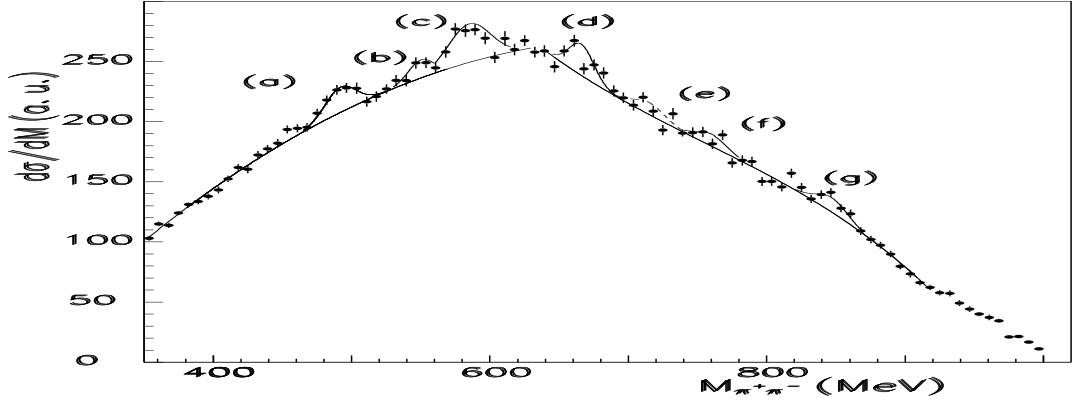}}
\caption{(Color on line). $M_{\pi^{+}\pi^{-}}$ invariant mass from the ${\it pp} \to {\it pp}{\it \pi^{+}\pi^{-}}$ reaction, Saturne (Disto beam line) \cite{balestra}.}
\label{fig:fig2}
\end{figure}
The ${\it pp} \to {\it pp\pi^{+}\pi^{-}}$ reaction was studied at Saturne (Disto beam line) \cite{balestra} using proton momenta $p_{p}$ = 3.67~GeV/c. Fig. 2 shows the $d\sigma/dM$ spectra  versus  
$M_{\pi^{+}\pi^{-}}$ invariant mass. The extracted peaks correspond to the following masses M $\approx$ 488[484.7] S.D. = 5.7, 550[549.7]
S.D. = 2.5, 585[584.7] S.D. = 6.4, 670[675.2] S.D. = 4.4. The masses previously observed in the SPES3 data are given between brackets. Small peaks poorly defined but appeared before in other data correspond to M $\approx$ 715[700], 760[754.7], and 847~MeV. 
\subsection{The ${\it pp} \to {\it pp}X^{0}$ reaction}
\begin{figure}[h]
\vspace*{0.cm}
\hspace*{-0.1cm}
\scalebox{0.97}[0.55]{
\includegraphics[bb=170 347 411 776,clip,scale=0.9] {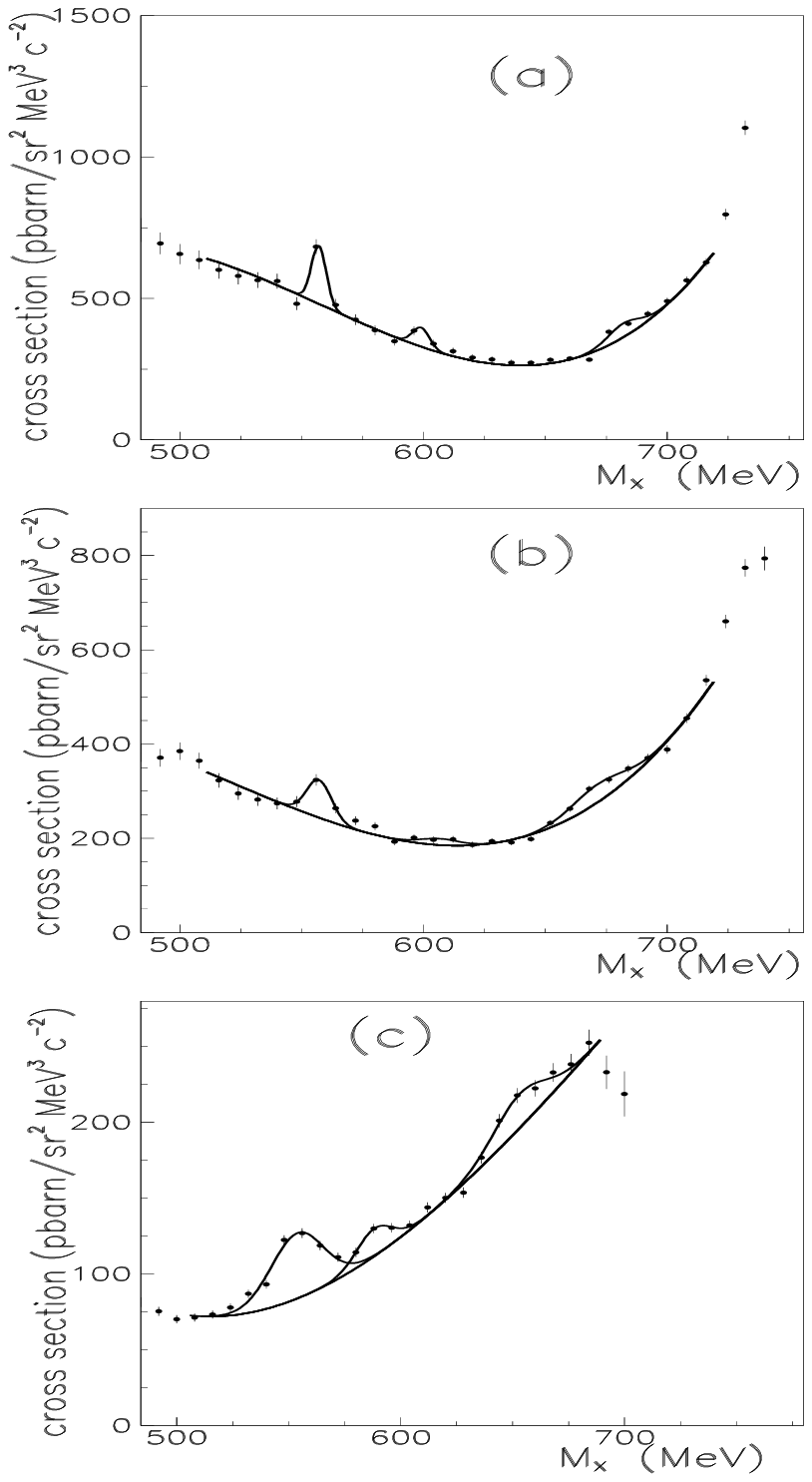}}
\caption{Cross sections of the ${\it pp} \to {\it pp}X^{0}$ reaction Saturne (SPES3) \cite{bormes}.}
\end{figure}
\begin{figure}[h]
\vspace*{-9.2cm}
\hspace*{7.6cm}
\scalebox{0.97}[0.56]{
\includegraphics[bb=181 349 413 774,clip,scale=0.9] {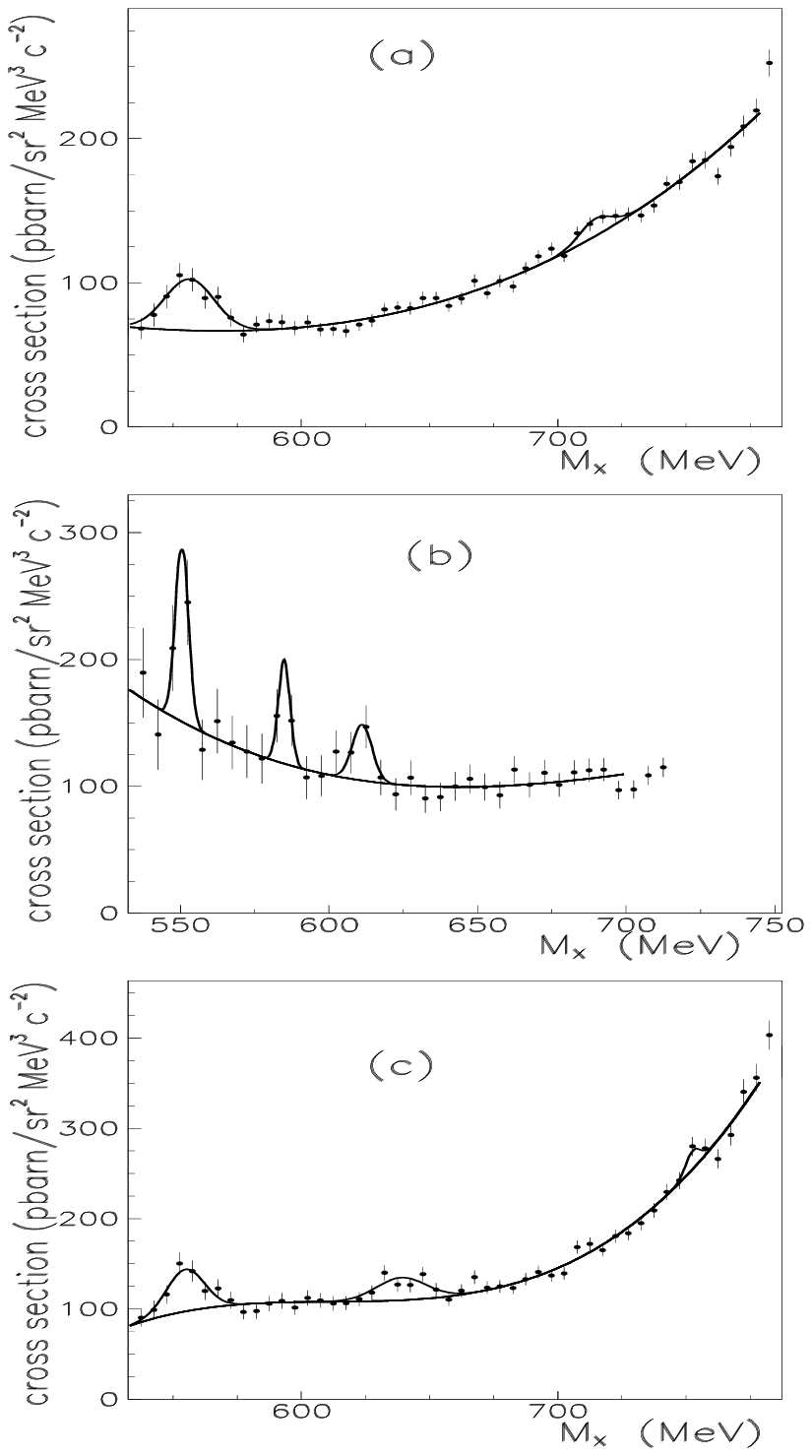}}
\end{figure}
\vspace*{0.9cm}

The ${\it pp} \to {\it pp}X^{0}$ reaction was studied at Saturne (SPES3 beam line) \cite{bormes}. Missing mass spectra $M_{X}$  were obtained at fifteen different experimental conditions, namely three incident proton energies: $T_{p}$ =  1520, 1805, and 2100~MeV, and several spectrometer angles at each energy.
Both protons were simultaneously detected in the same detection, thanks to its large momentum range 600$\le$pc$\le$1400~MeV/c. Peaks were extracted at the following masses: M = 62, 80, 100, 181, 198, 215, 228, 310, 350, 426, 470, 555, 588, 647, 681, 700, and 750~MeV. A selection of data is shown in Fig.~3; the S.D. of the peaks from left (l) to right (r) are in  (a)l:  2.9, 2.1, and 3.9; (b)l: 5.2, 2.9, and 6.1;
(c)l: 13.8, 7.7, and 2.5; (a)r: 5.0 and 2.3; (b)r: 1.0, 1.4, and 1.9; and  (c)r: 4.2, 2.7, and 1.7.
\subsection{The $e^{+}e^{-} \to \pi^{+}\pi^{-}$ reaction}
\begin{figure}[h]
\hspace*{-0.2cm}
\vspace*{1.0cm}
\scalebox{1.1}[0.5]{
\includegraphics[bb=36 139 527 546,clip,scale=0.8] {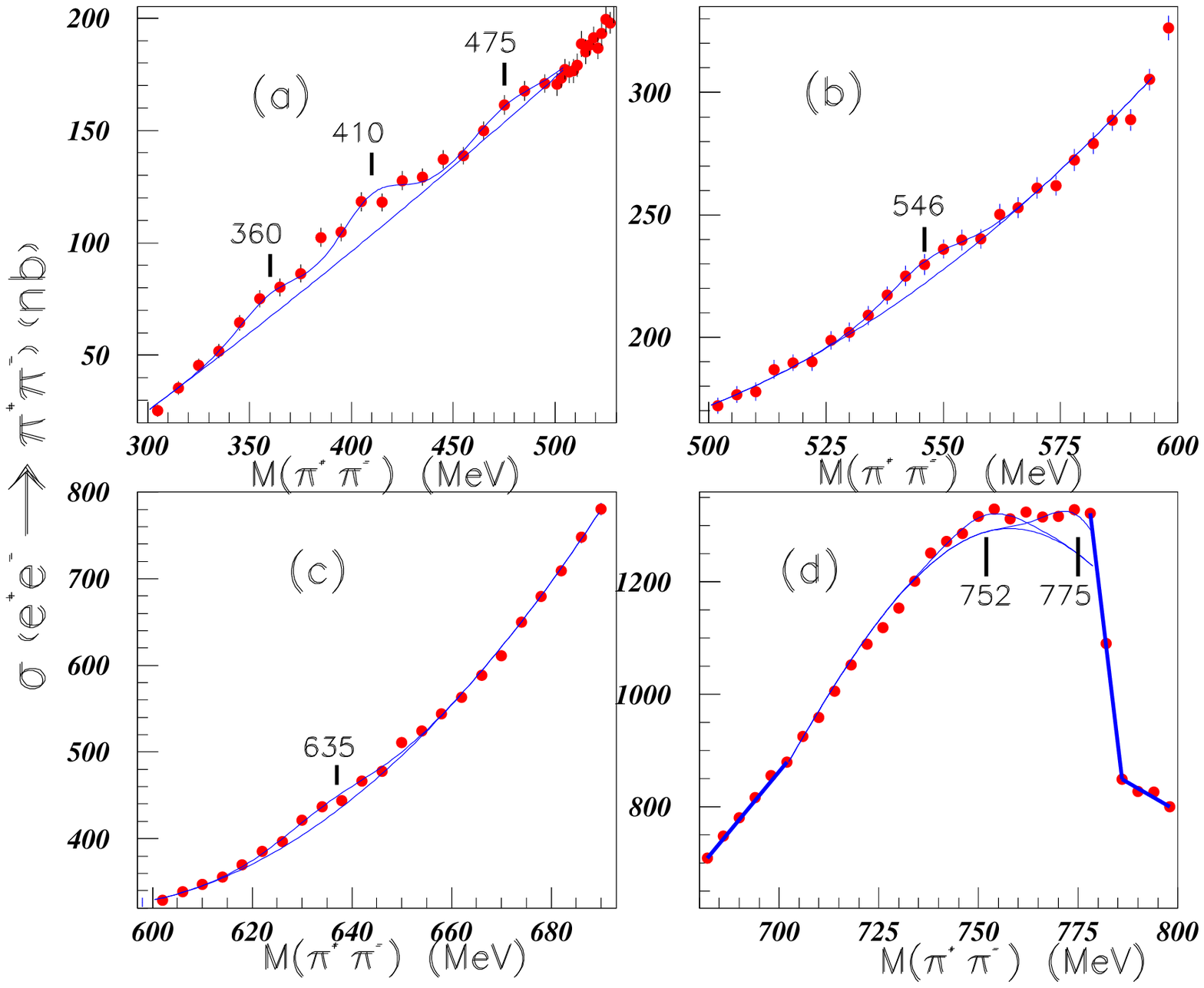}}
\caption{The $e^{+}e^{-} \to \pi^{+}\pi^{-}$ reaction, BABAR \cite{aubert}}
\end{figure}
The cross section of the $e^{+}e^{-} \to \pi^{+}\pi^{-}$ reaction was studied at BABAR  \cite{aubert}. Although small statistics are expected, since there is no quark in initial state, precise data with small binning were obtained at this electron-positron collider.  Fig.~4 shows several peaks extracted in the invariant $M_{\pi^{+}\pi{-}}$ spectra. They are shown with a 10~MeV binning in insert (a) and 4~MeV binning in the other three inserts. The masses of the extracted peaks and the corresponding statistical precisions are: insert (a) M = 360~MeV (S.D. = 3.1), 410~MeV (S.D. = 4.7), 475~MeV (S.D. = 1.4). Insert (b) M = 546~MeV (S.D. = 2.6). Insert (c) M=635~MeV (S.D. = 4.2). Insert (d) M = 752~MeV (S.D. = 3.4).
\section{Short recall of selected results already shown in previous papers}
This section recalls several data useful for the present study, using spectra showing narrow mesonic structure(s) in experiments studied for different aims. The extracted structures, have not been commented by the authors. They are mostly statistically not significant, i.e., the S.D. are small, therefore not always calculated. As already said, their justification lies in the existence of several such observations at the same (or nearby) mass. Table 2 shows some of these results. Some of these data and other old results are described shortly in \cite{bormes}. Among them the four two photon invariant masses measured from reactions involving high energy heavy ions \cite{hehi}. 
\begin{table}[h]
\begin{center}
\label{table:table2}
\caption{Recall of several narrow mesonic masses (in MeV), observed in various experiments.See text.}

\begin{tabular}[h]{c c c c}
\hline
observ.&reaction&lab.&ref.\\
\hline
$M_{\gamma \gamma}$&${\it p}{\bar p}$ ann. in flight&LEAR(Crystal Ball)& [9] \\
$M_{\gamma\gamma}$&$\eta \to \pi^{0}\gamma \gamma$&AGS(Crystal Ball)& [10]\\
$M_{\pi^{0}\gamma\gamma}$&$\eta \to \pi^{0}\gamma \gamma$&Mainz(Microtron)& [11]\\
$M_{X}$&$pp \to ppX$&Uppsala& [12]\\
M$_{\gamma\gamma}$&$dC \to \gamma \gamma$X&JINR(Nuclotron)& [13]\\
M$_{\pi^{+}\pi^{-}}$&$\gamma p \to p \pi^{+} \pi^{-}$&Desy& [14]\\
M$_{\pi^{-}\pi^{0}}$&$\gamma n \to p \pi^{-} \pi_{0}$&Mainz(MAMI)& [15]\\
M$_{\pi^{+}\pi^{0}}$&$\gamma p \to p \pi^{+} \pi^{0}$&Mainz(MAMI)& [16]\\ 
M$_{\pi^{0}\pi^{0}}$&pp$\to$pp$\pi^{0}$&Uppsala(Celsius)& [17]\\
M$_{X}$&ep$\to$e'pX&MAMI& [18]\\ 
M$_{X}$&ep$\to$e'pX&JLAB(Hall A)& [19]\\
M$_{X}$&ep$\to$e'pX&JLAB(Hall C)& [20]\\
M$_{\pi^{+}\pi^{-}}$&$pp\to pp\pi^{+}\pi^{-}$&Celsius& [21]\\
M$_{\gamma\gamma}$&$pp \to pp\gamma\gamma$&Celsius& [22]\\
\hline
\end{tabular}
\end{center}
\end{table}
\subsection{The ${\it p}{\bar p}$ annihilation in flight}
Fig. 2(a) of \cite{abele} shows the spectra of $\pi^{0}\gamma$  invariant mass using 1940~MeV/c proton momenta at LEAR by the Crystal Barrel Collaboration. The mass of the extracted structure M = 587~MeV, with a large S.D. = 7.1, is very close to M = 585~MeV, where a narrow peak was observed in the SPES3 data. 
\subsection{The $M_{\gamma\gamma}$ invariant mass}
The $\pi^{0}\gamma\gamma$ invariant mass from the rare $\eta $ decay, was studied at AGS with the Crystal Ball/TAPS \cite{prakhov} and 716~MeV/c $\pi^{-}$ beam.  In the M$_{\gamma\gamma}$ invariant mass, structures at M = 255[251], and 343[349]~MeV  with a low number of S.D.  are extracted. Although these extracted structures are rather insufficiently defined, they appear at masses that are close to masses of previously observed structures (in brackets). The additional peaks at M=84[81.3] and 190[194]~MeV, although being located close to already observed masses can, with a rather large certainty, be an artifact of the pion subtraction, and therefore they are not kept. 
\subsection{The $M_{\pi^{0}\gamma\gamma}$ invariant mass}
The doubly radiative decay $\eta \to \pi^{0}\gamma\gamma$ was studied at the Mainz Microtron MAMI with the Crystal Ball and TAPS multiphoton spectrometer. Fig. 7(d)  of Ref. \cite{nefkens} shows the results versus the $M_{\pi^{0}\gamma\gamma}$ invariant mass. Statistically poor structures are extracted at M $\approx$369[367], 415[415] and 479[482]~MeV.
\subsection{The M$_{X}$ missing mass of the $pp \to ppX$ reaction}
The $pp \to ppX$ reaction was studied at Uppsala \cite{wilkin} with the PROMICE-WASA facility. The missing mass obtained using the 310~MeV proton beam energy exhibits two structures at M = 61[61.5] and 76[81]~MeV.  
\subsection{The M$_{\gamma\gamma}$ invariant mass from the d C $\to \gamma\gamma$ X reaction}
The d C $\to \gamma\gamma$ X reaction was studied, with the internal 2.75~GeV/c per nucleon beam of the JINR Nuclotron \cite{abraamyan}. A structure at M$_{\gamma\gamma}$ = 360$\pm$7~MeV [366.7] was observed. It is noteworthy that these masses agree with the masses already obtained by other experiments.
\subsection{The $M_{\pi^{+}\pi^{-}}$ invariant masses from the study of the ${\it np} \to {\it np\pi^{+}\pi^{-}}$ reaction}
As already mentioned, the study of $M_{\pi^{+}\pi^{-}}$ invariant masses, started long time ago, with the analysis of bubble chamber, from the reaction ${\it np} \to {\it np\pi^{+}\pi^{-}}$  using $p_{n}$ = 5.20~GeV/c neutron beam  \cite{troyan}. Their statistics was progressively improved, leading to the following masses in the mass range studied here: M = 350$\pm$3[347] (S.D. = 3.0), 408$\pm$3[415] (S.D. = 3.5), 489$\pm$3[482] (S.D. = 4.0), 579$\pm$5[415] (S.D. = 3.8), 676$\pm$7[675.2] (S.D. = 3.0), and 762$\pm$11[754.7] (S.D. = 6.1).
\subsection{The $M_{X}$ missing  mass from the pp$\to pp\pi$X reaction}
The two-pion production in p-p scattering was studied with the WASA/PROMICE detector at the CELSIUS storage ring.  Fig. 5 of Ref. \cite{bilger} shows the missing mass $M_{X}$ of the $pp \to pp\pi$X reaction. Two structures close to 40 and 80~MeV are observed when the pion is identified using only $\Delta$E/E information as well as a structure close to 35~MeV when a delayed pulse from the pion is requested in addition.
\section{Careful scrutiny of previous data}
\begin{figure}[h] 
\scalebox{0.85}[0.8]{
\hspace*{-0.28cm}
\includegraphics[bb=1 45 515 544,clip,scale=1] {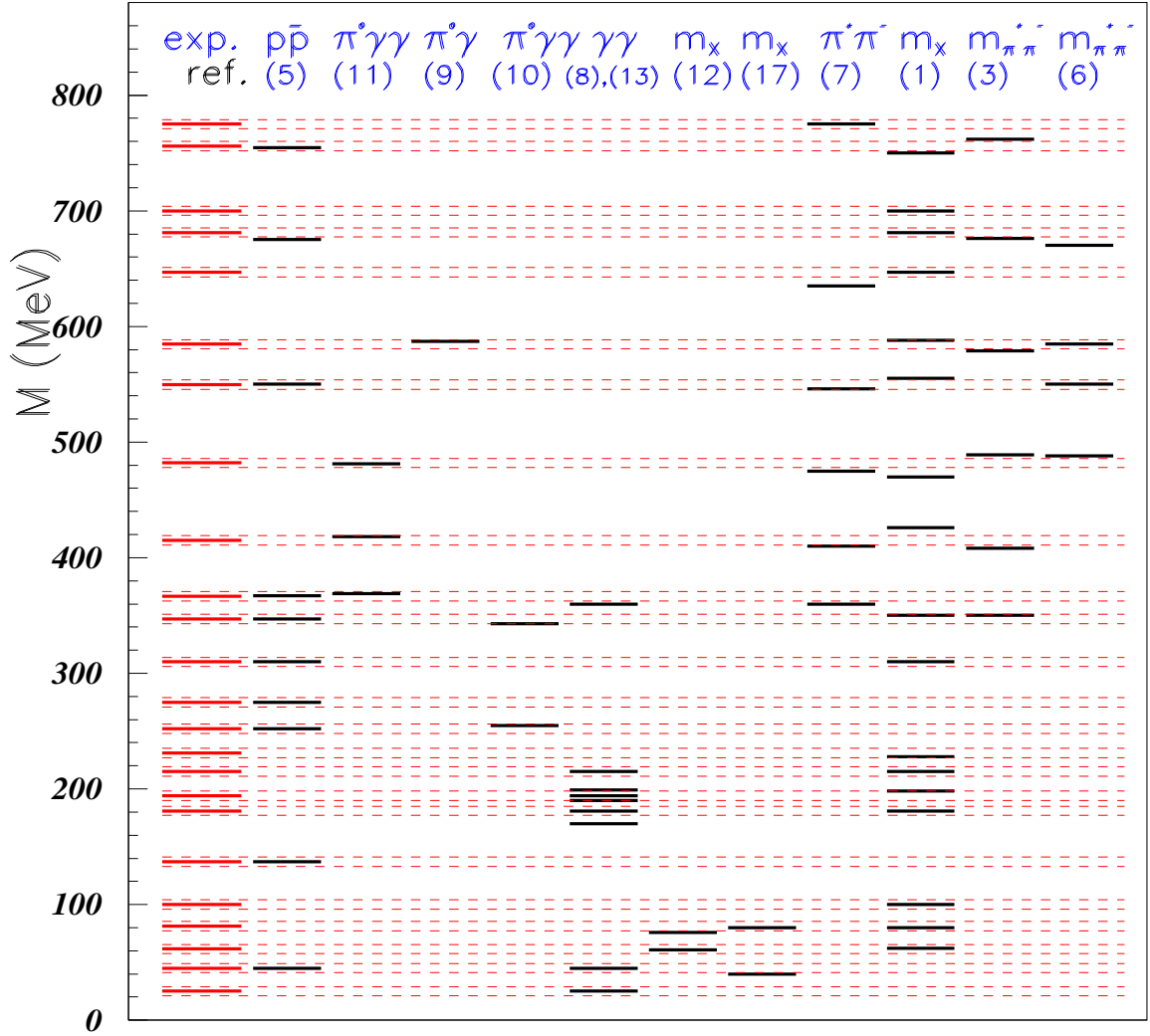}}
\caption{Masses (in MeV) of narrow unflavoured mesonic structures extracted from several different experimental data.}
\end{figure}

The two masses from SPES3, at M = 227~MeV and 235~MeV, are merged by a common one at M = 231~MeV.

All masses quoted above are presented in Fig. 5, with the attempt to attribute the same rank to nearby masses from different experiments. Structures extracted from different reactions are often observed  at nearby masses. The mean values of these masses  are: M = 25, 45.3, 61.5, 81.3, 100, 181, 194, 215, 231, 252, 274.9, 309.9, 347, 366.7, 415, 482, 549.7, 584.7, 641, 675.2, 700, and 754.7~MeV.
\section{$\pi-\pi$ phase-shifts}
Three types of experiments studied the pion-pion phase-shifts. The first
one consists to perform and analyze the $\pi^{-}$p experiments
producing two-pion and a neutron final states, either $\pi^{0}\pi^{0}$n,
either $\pi^{+}\pi^{-}$n. Since low energy pion beams are difficult to
use, and the rescattering from final neutron is present, the analysis is
model-dependent, and the phase-shifts at low
two-pion masses are scarce and not precise. For example below
M$_{\pi\pi}$ = 450~MeV, the values of $\delta^{0}_{0}$ measured with
2~GeV/c  or 7~GeV/c \cite{maen} incident pion momentum are
very different.\\
\hspace*{4.mm}
The second type of experiments consist of studying the $\bar{N}N$ annihilation
in flight into two nucleons. However the low energy data, at $T_{p}$ = 66.7~MeV,
produce pions which C.M. energy is as high as 815~MeV. Then these
experiments are not suitable for low energy phase shift determination.
\hspace*{4.mm}
The third type of experiments, called K$_{e4}$, measures and
analyses in a model-independent way, the K$^{+}$ disintegration:
K$^{+}\to\pi^{+}\pi^{-}e^{+}\nu$ \cite{ross} \cite{pisl}, or
K$^{+}\to\pi^{0}\pi^{0}e^{+}\nu$ \cite{shim}. In these experiments, the
branching ratios are small; however, these experiments are often used
since the two pions are the
only hadrons in the final state. The scattering lengths and
principally  the isoscalar S-wave scattering length a$^{0}_{0}$ are
measured (and calculated \cite{cola}), since they are connected
to the chiral perturbation theory. But the isospin 2 phase-shift
is poorly known \cite{hoog} and so are the imaginary parts
of these phase-shifts (which should exist at all M$_{\pi\pi}$
masses since the charge exchange $\pi^{+}\pi^{-}\to\pi^{0}\pi^{0}$
can occur). Finally the phase-shifts $\delta_{l}^{I}$ and the
inelasticities $\eta_{l}^{I}$ are not known between 400~MeV and
600~MeV \cite{kami,kloe,gray,hyam}. There is
a large difference, in the range 1.0$\le\sqrt{s}\le$1.3~GeV between
$\eta_{0}^{0}$(s) determined from $\pi\pi\to\pi\pi$ and from
$\pi\pi\to\bar{K}K$ \cite{pela}. All three inelasticities:
$\eta_{0}^{0}$, $\eta_{0}^{2}$, and $\eta_{2}^{0}$ are set to one 
(no inelasticity) for s$\le$1~GeV in this work \cite{pela}.

There is no room, in the present pion-pion phase shifts,
for signatures of the narrow structures shown in the previous
sections of the present work. Narrow resonances are easy to miss
in phase-shifts, all the more they are narrow and weakly excited.

It is clear that our results will slightly modify the low
energy pion-pion phase-shifts. It is also clearly outside the
scope of the present work to give quantitative information on
the consequence of our observations on the low energy pion-pion
phase-shifts.

\section{Recall of data showing the presence of narrow low mass unflavoured baryons}
\normalsize
\begin{figure}[ht]
\vspace*{0.0cm}
\hspace*{0.0cm}
\scalebox{1}[1]{
\includegraphics[bb=55 613 544 795,clip,scale=0.9]{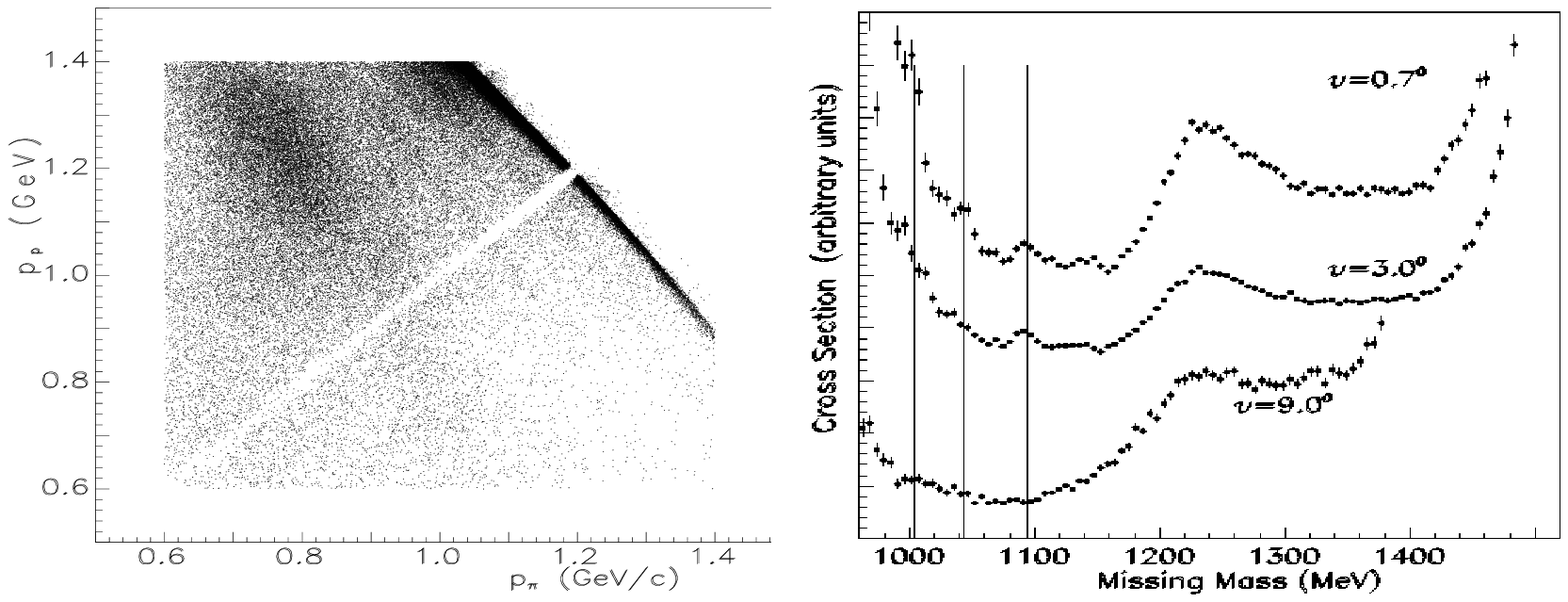}}
\caption{Scatterplot of $p_{p}$ versus $p_{\pi^{+}}$ at T$_{p}$ = 1805~MeV and $\theta$ = 0.75$^{0}$ (left), and the corresponding $M_{X}$ spectra (right) \cite{bormes}.}
\end{figure}
\normalsize

The experimental evidence for narrow low mass baryons was already observed thirty years ago \cite{borbar}. The masses of the narrow corresponding structures are: M = 1004, 1044, 1094, 1136, 1173, 1249, 1277, and 1384~MeV. They were observed in the missing mass $M_{X}$ and in the invariant masses: $M_{p\pi^{+}}$ and $M_{\pi^{+}X}$ of the pp$\to$p$\pi^{+}$X reaction studied at SPES3 (Saturne). Later on, these studies were extended to the mass range 1470$\le$M$\le$1680~MeV \cite{borbar1} and to the mass range 1720$\le$M$\le$1790~MeV
\cite{borbar2}. The study of the mass range 1470$\le$M$\le$1680~MeV, allows us to make the conjecture that the broad Particle Data Group (PDG) baryonic resonances: N(1520)D$_{13}$, N(1535)S$_{11}$, $\Delta$(1600)P$_{33}$, N(1650)S$_{11}$, and N(1675)D$_{15}$, are collective states built from several narrow and weakly excited resonances, each having a (much) smaller width than the one reported by PDG.

Fig. 6 (left) illustrates the data at T$_{p}$ = 1805~MeV and $\theta$ = 0.75$^{0}$.\\
\begin{figure}[ht]
\hspace*{-0.5cm}
\centering
\scalebox{0.9}[0.5]{
\includegraphics[scale=0.95]{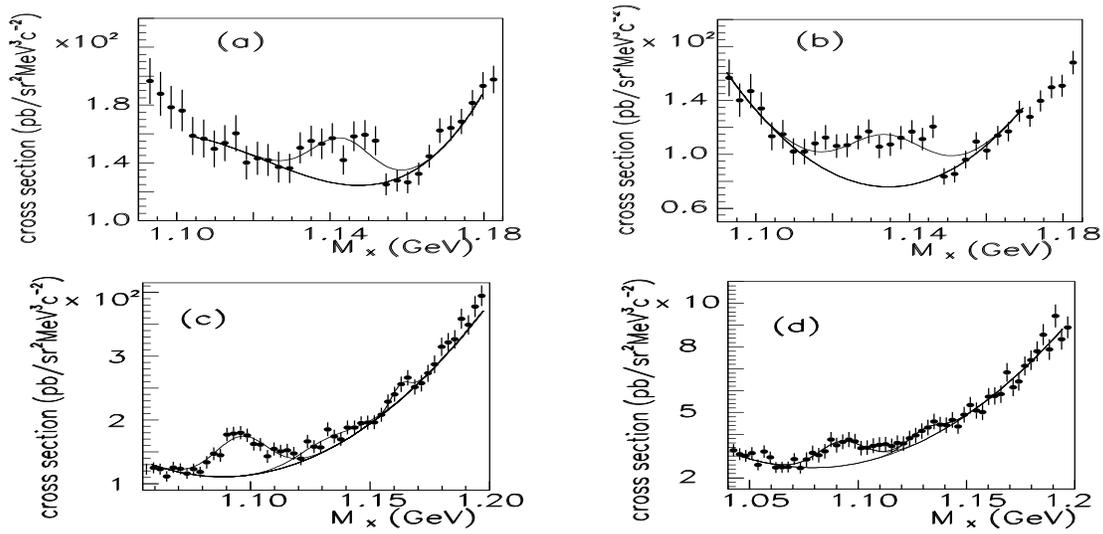}}
\caption{Selection of several M$_{X}$ spectra (see text).}
\end{figure}
 Fig. 7 illustrates several spectra, namely:\\
 insert (a), T$_{p}$ = 2.1~GeV, $\theta$ = 0.7$^{0}$, M = 1143~MeV, S.D. = 6.3\\
 insert (b), T$_{p}$ = 2.1~GeV, $\theta$ = 3$^{0}$, M = 1133.7~MeV, S.D. = 11\\
 insert (c), T$_{p}$ = 1.805~GeV, $\theta$ = 3.7$^{0}$, M = 1130.4~MeV, S.D. = 4.9\\
 insert (d), T$_{p}$ = 1.905~GeV, $\theta$ = 9$^{0}$, M = 1130.8~MeV, S.D. = 3.6. 

The mass range 946$\le$M$\le$995~MeV was also studied through the same reaction as used for the study of mass range above 1000~MeV \cite{btdou02}. Several masses were tentatively extracted, all from a small number of data. Two masses only are observed in more than two spectra, namely at M = 973 and 986~MeV. The same mass range was studied at the proton Linear Accelerator of INR \cite{filkov} with help of the $pd \to ppX_{1}$ reaction. Three narrow masses were observed at $M_{X_{1}}$ = 966$\pm$2, 986$\pm$2, and 1003 $\pm$ 2~MeV. The two larger masses are the same as those observed in the Spes3 data. The first one differs by 7~MeV.  Since the precision in Ref. \cite{filkov} is better, the M = 966~MeV value is kept.
\subsection{The $p(\alpha,\alpha')X$ reaction}
The  $p(\alpha,\alpha')X$ reaction was studied at Saturne(SPES4) \cite{morsch} in order to study the radial excitation of the nucleon in the $P_{11}$(1440) Roper resonance. The data were collected at the incident beam energy $T_{\alpha}$ = 4.2~GeV and two angles $\theta$ = 0.8$^{0}$ and 2$^{0}$. \\
\begin{figure}[ht]
\vspace*{0.0cm}
\hspace*{0.0cm}
\scalebox{1}[1]{
\includegraphics[bb=74 588 500 795,clip,scale=0.95]{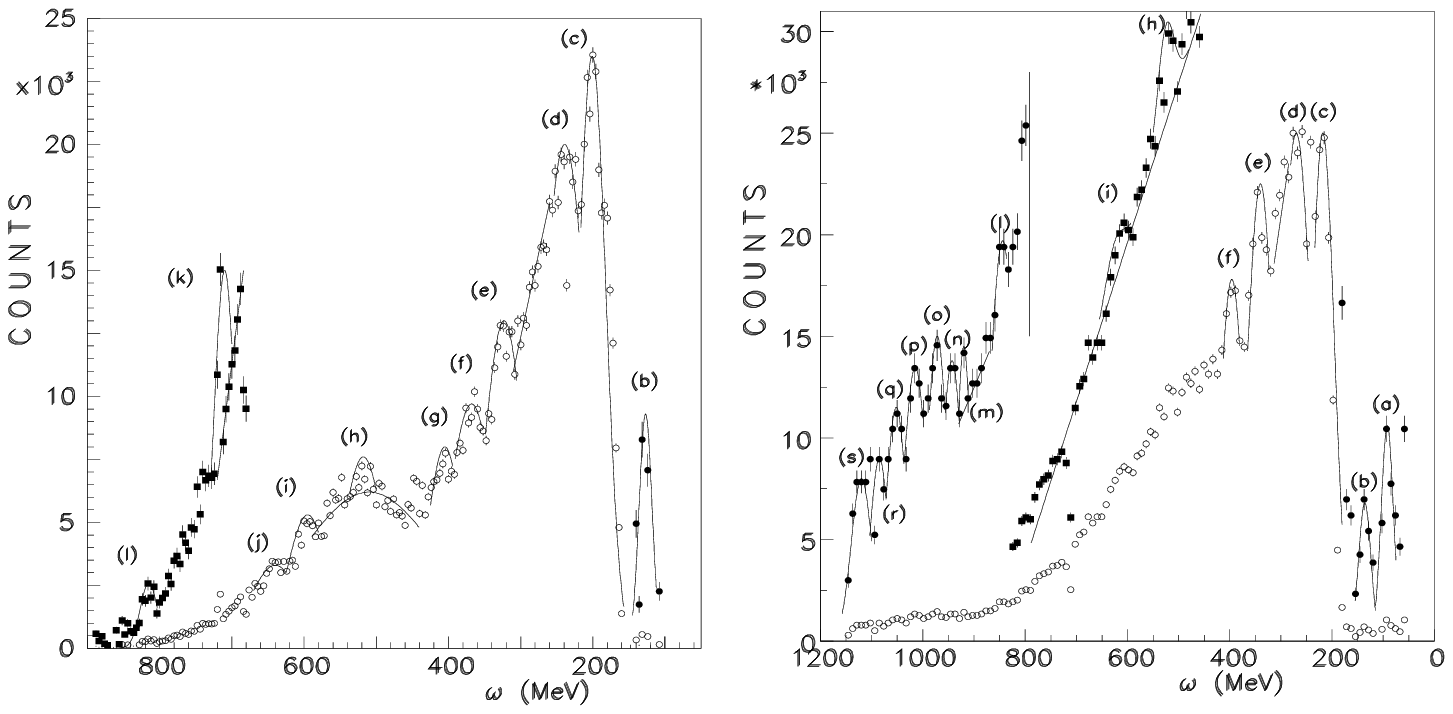}}
\caption{ Cross-section of the p($\alpha,\alpha'$)X reaction at $T_{\alpha}$ = 4.2~GeV, $\theta$ = 0.8$^{0}$ (left), and 2$^{0}$ (right) \cite{morsch}.}
\end{figure}
\begin{figure}[ht]
\vspace*{0.7cm}
\hspace*{0.5cm}
\scalebox{0.8}[0.5]{
\includegraphics[bb=20 18 523 521,clip,scale=0.9]{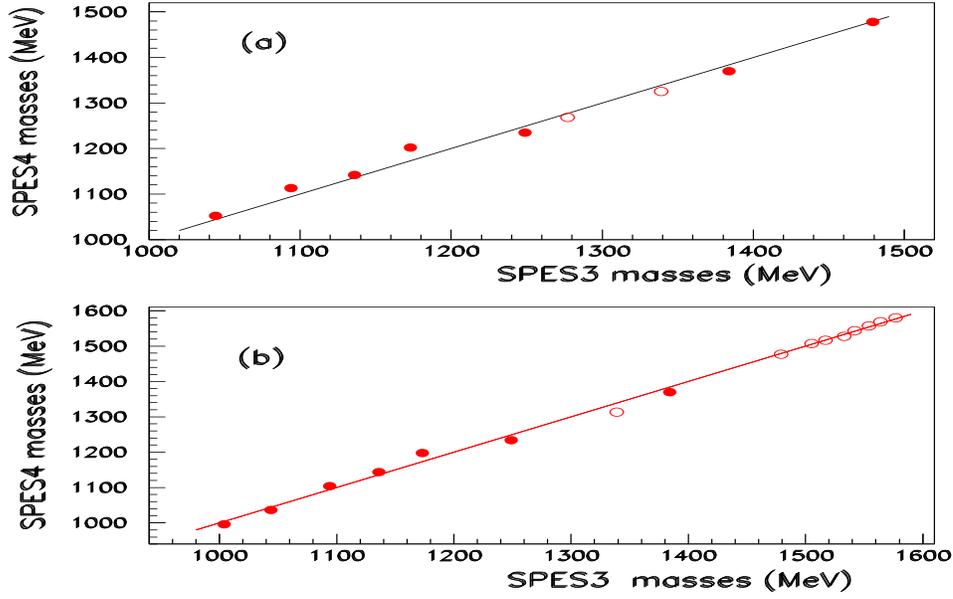}}
\caption{(Color on line). Comparison between masses of narrow baryons extracted from SPES3 and SPES4 \cite{morsch} data. Insert (a) and (b) correspond respectively to $\theta$=0.8$^{0}$ and $\theta$=2$^{0}$.}
\end{figure}
Above projectile and target excitations, the histograms taken at both energies exhibit well statistically defined structures with large S.D. values. Their masses agree with the masses observed in the SPES3 data as seen in Fig. 9 \cite{var06}.
\subsection{The virtual Compton scattering cross-sections}
The virtual Compton scattering cross-sections were measured by the Hall A Collaboration at JLAB \cite{laveissiere}. Fig. 6 of \cite{var09} shows a good fit when the spectra are analyzed including the masses of structures from SPES3. All peaks here have the same width (FWHM = 36~MeV). We observe the need to introduce an additional mass at M=1212~MeV.
\subsection{The p(d,2p)$\Delta^{0}$ reaction}
The cross section of the p(d,2p)$\Delta^{0}$ reaction was measured at Saturne(SPES4) using 2.0 and 1.6~GeV incident deuterons at several spectrometer angles \cite{ellegarde}. They exhibit oscillating behaviours in the low part of the $\Delta$ missing mass range. The corresponding structure masses agree with the masses observed previously in different data \cite{var09}.
\subsection{Narrow structures observed in charge exchange reactions}
Several reactions of charge exchange reactions, were reported in \cite{var09}, where narrow structures lie above continuous spectra. They are only quoted here, the corresponding references and quantitative discussions are given in \cite{var09}. In addition to the reaction quoted just above, we note the following reactions mainly measured at Saturne (SPES4) \cite{ellegarde}:
\begin{itemize}
\item p(d)($^{4}He,t)\Delta^{++}$(X)  reactions measured at Saturne (SPES1) and (SPES4),
\item p($^{12}C,^{12}N)\Delta^{0}$ reactions measured at Saturne (SPES4),
\item p($^{20}Ne,^{20}Na)\Delta^{0}$ reactions measured at Saturne (SPES4).
\end{itemize}
as well as similar reactions on light heavy ion targets as $^{12}C$ and $^{27}Al$.

Let us also mention systematic studies on p(p,n)$\Delta^{++}$ measured at LAMPF.

Several spectra, concerning narrow baryonic as well as narrow dibaryonic structures were also published in several papers \cite{btjy}. The mean value of the masses kept as masses of narrow exotic baryons are obtained, as was done for narrow mesons, by giving the same rank for close mass structures. They are: M = 966, 986, 1004, 1094, 1136, 1173, 1210, 1249, 1277, 1339, 1384, 1423, and 1480~MeV.\\

The effect of the narrow baryonic structures on NN phase shifts, should be very small, since their excitation is much smaller than the PDG mass excitations. So the ratio of M(1004)/M(939.6) is $\approx$ 1.6 $10^{-3}$ at 
$T_{p}$ = 1805~MeV and $\theta$ = 3.7$^{0}$ and decreases for increasing angles. The corresponding quantitative study of the N-N phase shifts is clearly outside the scope of the present study. 
\section{Recall of data showing the presence of narrow low mass unflavoured dibaryons}
The evidence for narrow low mass dibaryons was studied 27 years ago \cite{bordibar1}. 
They were observed in the missing mass spectra of the $^{3}$He(p,d)X reaction measured at Saturne, at the SPES1 beam line, with a proton beam $T_{p}$ = 1.2~GeV, at $\theta_{d}$ = 33$^{0}$.
Narrow structures were observed at $M_{X}$ = 2.122, 2.198, and possibly at 2.233~GeV.  A similar experiment was performed at Lamp \cite{santi}. The study of the analyzing power of the 
$^{3}He(\vec{p},d)X$ reaction. allowed to observed narrow structures at M = 2.015$\pm$0.005, 2.054$\pm$0.006, 2.125$\pm$0.003, 2.152$\pm$0.004, and 2.181$\pm$0.005~MeV.

The narrow unflavoured dibaryons were also observed in the invariant mass M$_{pX}$ of the already mentioned pp$\to$p$\pi^{+}$X reaction, studied at SPES3 (Saturne) \cite{bordibar}. The first extracted masses, are: M = 2050, 2122, and 2150~MeV. 

In the invariant mass $M_{pX_{1}}$ of the pd$\to ppX_{ 1}$ reaction, also mentioned before \cite{filkov}, three narrow peaks were observed at M = 1904$\pm$2, 1926$\pm$2, and 1942$\pm$2~MeV, all with an experimental width of 5~MeV. The first and third masses correspond exactly to masses previously  observed. The second mass 
M = 1926~MeV is very clearly observed in \cite{filkov}, and is therefore kept.

The $pp \to pp\gamma\gamma$ reaction was studied at the phasotron at JINR \cite{khrykin}. A narrow peak at one photon energy E = 24~MeV, was interpreted as a signature of an exotic dibaryon resonance with a mass M$\approx$1956$\pm$6~MeV (S.D. =5.3 $\sigma$) disintegrating into $pp\gamma$.

Several other masses, quoted in \cite{bordibar},  were observed, allowing to summarize the narrow, low mass, dibaryon masses at the following values: M = 1902, 1916, 1941, 1956, 1969, 2016, 2052, 2087, 2122, 2155, 2194, 2236, and 2282~MeV. 
\section{Discussion}
\subsection{Comparison between successive low mass exotic hadrons}
The previously reported narrow mesonic masses are shown in Fig. 10.
\begin{figure}[ht]
\vspace*{0.7cm}
\hspace*{0.9cm}
\scalebox{0.9}[0.7]{
\includegraphics[bb=1 47 488 516,clip,scale=0.9]{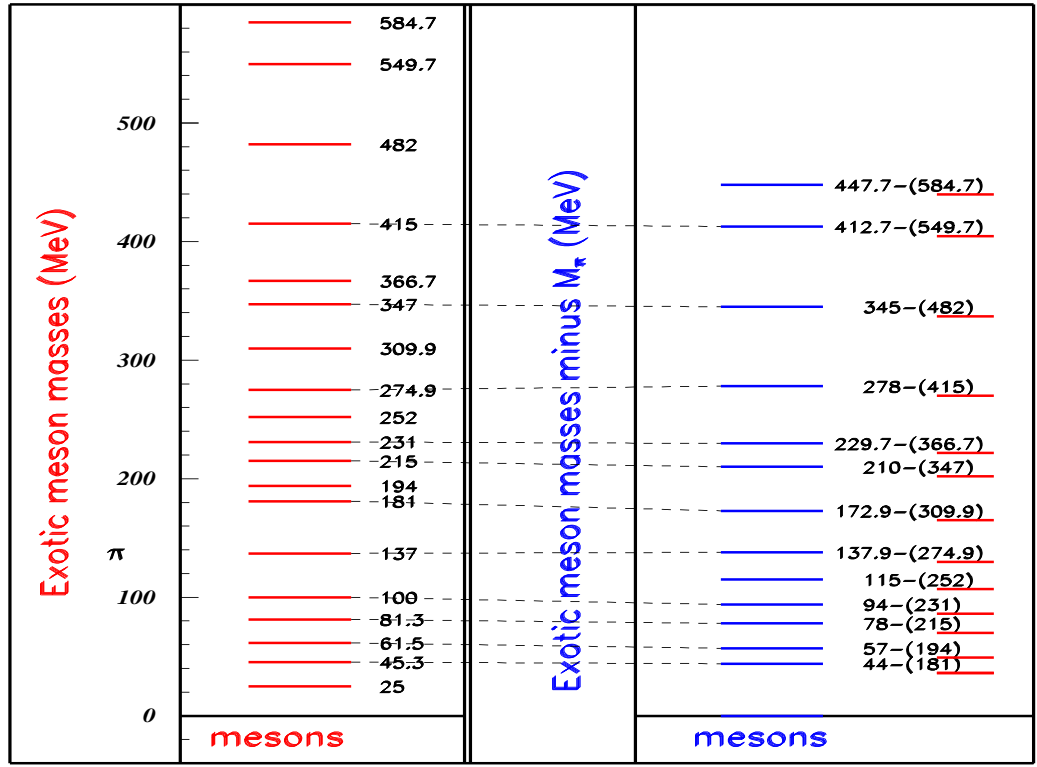}}
\caption{(Color on line). Masses of narrow exotic mesons (left) and the same (right) after pion mass subtraction.}
\end{figure}
After subtraction of the pion mass, the right part of this figure shows a nice correspondence of masses larger than pion with the first  masses (lower than the pion) plus the pion mass. For masses larger than twice the pion mass, the exotic meson mass is close to two times the pion mass plus the low narrow meson masses. As an example,  M = 181~MeV $\approx$ M$_{\pi}$ + 45.5~MeV and M = 215~MeV $\approx$ M$_{\pi}$ + 81.3~MeV. This observation suggests that we observe a significant coupling between the pion mass and the low exotic masses. But we observe also similar couplings between two (M $\approx$ 275~MeV), three (M $\approx$ 415~MeV), or four (M $\approx$ 549.7~MeV) pion masses. Such nearness suggests that we observe molecular states. The low mass exotic mesons, below pion mass, behave like multiquark-antiquark balls, with intermediate time lives. The same coupling is supposed to exist with stable nuclei, but they cannot be observed, being lost among nuclear levels.

Since we deduce from these mesonic low mass states, a significant coupling with stable pion, we are naturally lead to look at the possible same coupling with stable baryons. Fig. 11 shows that is indeed the case. \\
\begin{figure}[ht]
\vspace*{0.7cm}
\hspace*{0.9cm}
\scalebox{0.88}[0.7]{
\includegraphics[bb=20 51 492 517,clip,scale=0.9]{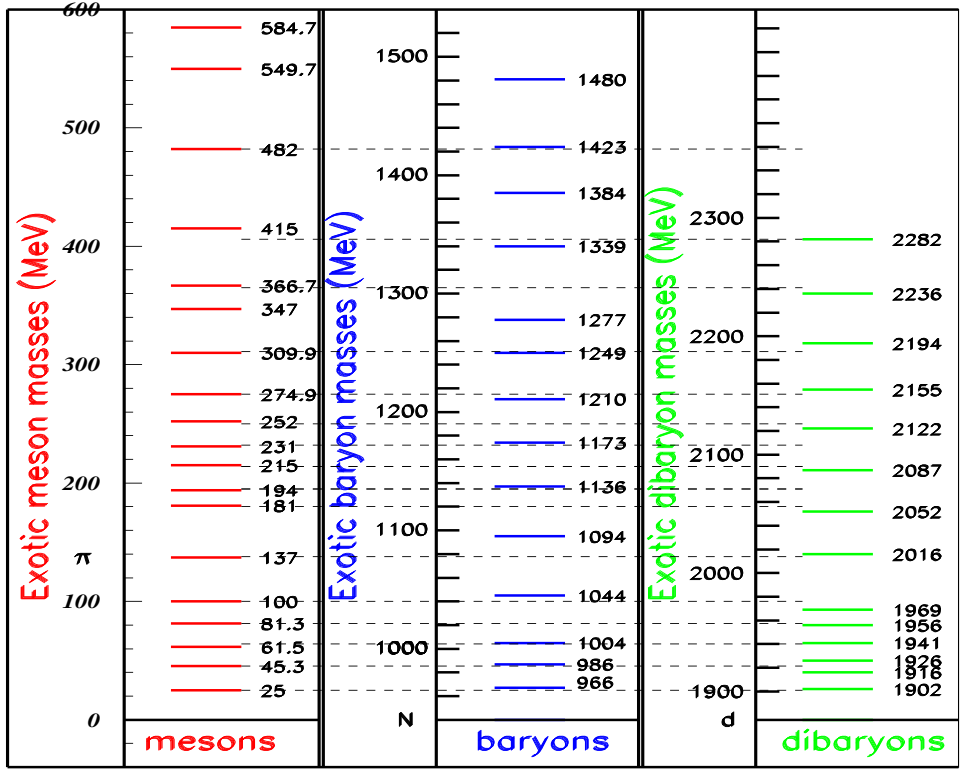}}
\caption{(Color on line). Masses of narrow exotic mesons, baryons and dibaryons with their own scale in one side, and compared to the narrow meson masses after subtraction of the stable corresponding mass: N or d. }
\end{figure}
In Fig. 11, the masses of exotic mesons, baryons, then dibaryons are plotted, each with its own scale, and the baryons minus nucleon mass, then dibaryons minus deuteron mass, are plotted in the scale of mesons. We observe a good correspondence up to the pion mass. But between M = 181~MeV and M = 252~MeV, we note that the mesons masses correspond alternatively to a baryon or a dibaryon mass. This property is not presently understood.
\subsection{Interpretation within two quark-antiquark clusters}
The observation presented above shows that it is sufficient to describe the mesonic masses. An attempt is described below in terms of two quark-antiquark clusters.  As already mentioned, the narrow mesonic structures discussed above, cannot be described within classical ${\it q}{\bar q}$. An attempt is therefore presented below to associate the masses of the mesonic structures with masses computed within a simple mass relation based on two quark clusters. A mass formula was derived some years ago for two clusters of quarks at the ends of a stretched bag in
terms of color magnetic interactions \cite{mul}.

A tentative description of the exotic low mass structures in mesonic, baryonic, and dibaryonic spectra, using this mass formula in a phenomenological approach, has been previously proposed. The mesonic structure masses were described, using $q^2- \bar q^2$, or $q^3- \bar q^3$ clusters up to M = 620~MeV, and $ q^4- \bar q^4$ between M = 620~MeV and M = 750~MeV
\cite{bormes}. The  narrow baryonic structure masses were described either by  $ q~-~ q^2$ or $ qqq~ - ~ q\bar q $ clusters
\cite{borbar}. Finally the dibaryonic structure masses were described \cite{bordibar} using the  $ q^2~-~ q^4$ configuration.                                                                                                                                                                                                                                                                                                                                                                                                                                                                                                                                                                                                                                                                                                                                                                                                                                                                                                                                                                                                                                                                                                                                                                                                                                                                                                                                                                                                                                                                                                                                                                                                                                                                                                                                                                                                                                                                                                                                                                                                                                                                                                                                                                                                                                                                                                                                                                                                                                                                                                                                                                                                                                                                                                                                                                                                                                                                                                                                                                                                                                                                                                                                                                                                                                                                                                                                                                                                                                                                                                                                                                                                                                                                                                                                                                                                                                                                                                                                                                                                                                                                                                                                                                                                                                                                                                                                                   
The following equation was used:                                                                                                                                                                                                                                                                                                                                                                                                                                                                                                                                                                                                                                                                                                                                                                                                                                                                                                                                                                                                                                                                                                                                                                                                                                                                                                                                                                                                                                                                                                                                                                                                                                                                                                                                                                                                                                                                                                                                                                                                                                                                                                                                                                                                                                                                                                                                                                                                                                                                                                                                                                                                                                                                                                                                                                                                                                                                                                                                                                                                                                                                                                                                                                                                                                                                                                                                                                                                                                                                                                                                                                                                                                                                                                                                                                                                                                                                                                                                                                                                                                                                                                                                                                                                                                \begin{equation}
M=M_0+M_1[i_1(i_1+1)+i_2(i_2+1)+(1/3)s_1(s_1+1) +(1/3)s_2(s_2+1)]
\label{eq:eq2}
\end{equation}
where $M_0$ and $M_1$ are parameters deduced from  experimental mass
spectra and $i_1(i_2)$, $s_1(s_2)$ are the isospin and spin of the first (second)
quark cluster.  
The same approach is employed here. Eq. (\ref{eq:eq2}) involves a
large degeneracy.
We made the assumption that the simplest configuration is
preferred, although additional $({\it q}{\bar q})^{2}$ terms allow to get the same results. The addition of  one 
${\it q}{\bar q}$ term allows to increase possible spins and isospins, but modifies also the parity.  

Eq. (\ref{eq:eq2})  is applied to two quark clusters $ q^n ~-~\bar q^n$. The simplest $ q - \bar q$ choice corresponds  to the most strongly excited meson, i.e., to the pion. Since therefore $ s_1=s_ 2= 
i_1=i_2=1/2$, the previous equation becomes $M = M_{0} + 2M_{1}$ = 137~MeV. The next cluster configuration is $ q^2  - \bar q^2$. The corresponding minimum mass is obtained with 
$s_1=s_2= i_1=i_2=0$.  Corresponding to these values, $M = M_{0}$. We associate this mass to our minimum narrow meson mass, therefore  $M_{0}$ = 25~MeV, from which one concludes that 
$M_{1}$ = 56~MeV.

\begin{figure}[ht]
\hspace*{0.6cm}
\scalebox{.9}[.52]{
\includegraphics[bb=28 256 284 544,clip,scale=0.83] {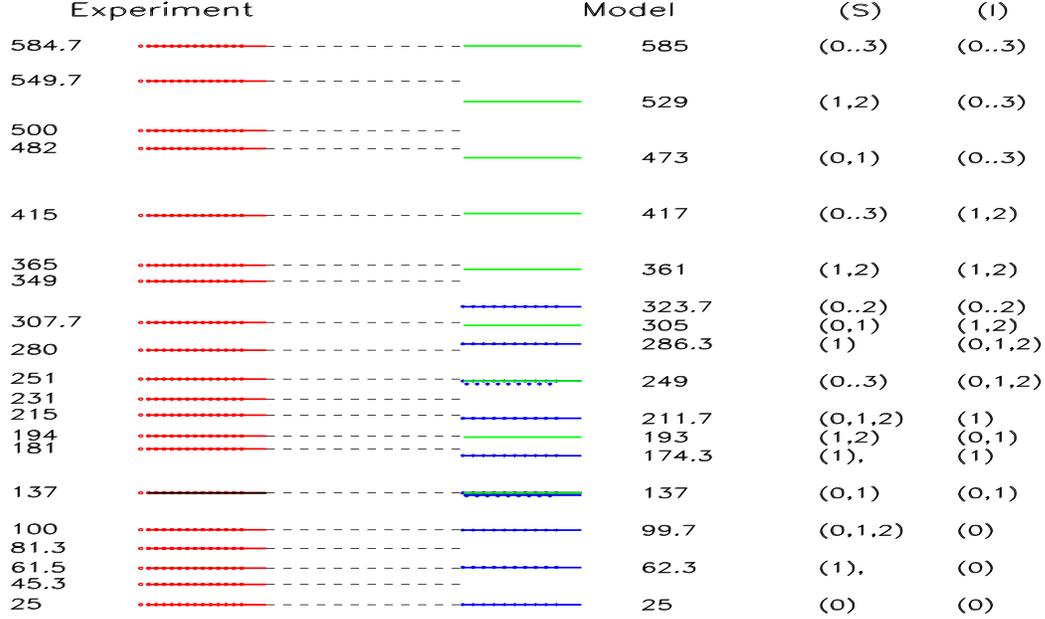}}
\vspace*{3.8cm}
\caption{(Color on line). The experimental exotic narrow meson masses are shown with red lines and empty circles. Calculated masses, using Eq. (1) 
(see text), are shown with blue lines and stars when obtained by ${\it q}^2  -{\bar q}^2$ clusters,
and shown with green lines when obtained with ${\it q}^3  - {\bar q}^3$ clusters.
 The possible spins and isospins corresponding to the predicted masses are also given.}
\label{fig:fig12}
\end{figure}

Experimental exotic narrow meson masses are shown in Fig.~12 with red lines overcomed by empty circles.
The masses are calculated with the previous $M_{0}$ and $M_{1}$ values, and are also shown in Fig. 12. 
The assumption for the $q^2 - \bar q^2$ quark clusters corresponds to masses drawn with blue lines and blue stars. The same masses are obtained within the assumption of 
$q\bar q -  q \bar q$ quark clusters. The assumption for  $q^3 - \bar q^3$ quark clusters, corresponds to masses drawn in green. All calculated masses using $q^2 - \bar q^2$ and $q^3 - \bar q^3$ quark clusters, are drawn in Fig. 12.

The figure is limited to M = 600~MeV, since with the $M_{0}$ and $M_{1}$ parameter values,  $s_1 = s_2 = 
 i_1 = i_2$ = 3/2 correspond to M = 585~MeV. 

Larger masses can be obtained with introduction of heavier quark clusters. It is noteworthy that the assumption $q^5\bar q^5$ quark clusters, allows to find several experimental masses. Table~\ref{tab:table3} shows these masses, limited to a value M$\le$780~MeV.
\begin{table}[h]
\begin{center}
\vspace*{-0.2cm}
\caption{Masses calculated within  $ q^4\bar q^4$ quark clusters, and some calculated masses within the 
$q^5\bar q^5$ quark clusters, compared to the experimental values.}
\label{tab:table3}
\vspace{0.mm}
\begin{tabular}{c c c c c c c}
\hline
conf. &$s_{1}$& $s_{2}$ &$i_{1}$&$ i_{2}$& M(MeV)& Exp.(MeV)\\
\hline
${\it q}^{4} - {\bar q}^{4}$&0 &1 &0 &2 &398.3& \\
                                   &1&1&0&2&435.7& \\
                                   &1&2&0&2&510.3& \\
                                   &1&1&1&2&547.7&549.7 \\
                                   &1&2&1&2&622.3&\\
                                   &2&2&1&2&697&700\\
                                   &0&1&2&2&734.3&\\
                                   &1&1&2&2&771.7&\\
                                   &0&2&2&2&803&\\
                                   &1&2&2&2&846.3&\\
                                   &2&2&2&2&921&\\
${\it q}^{5} - {\bar q}^{5}$& 3/2& 3/2 &3/2 &3/2 &585 &584.7\\
&1/2&1/2&1/2&5/2&585&584.7\\
&3/2&1/2&1/2&5/2&641&641\\
&5/2&3/2&3/2&3/2&678.3&675.2\\
&3/2&3/2&1/2&5/2&697&700\\
&1/2&1/2&3/2&5/2&753&754.7\\
\hline
\end{tabular}
\end{center}
\end{table}

The predicted possible spins and isospins are obtained from spins and isospins of both clusters.
The predicted parities for exotic mesons depend on the number of ${\bar q}$. The negative parity of the $\pi$ is obtained with 
${\it q} - {\bar q}$ or ${\it q^3} - {\bar q^3}$. The $ q^2 - \bar q^2$ choice necessitates an orbital excitation between both clusters to get the negative parity. 

All masses obtained from the ${\it q}^{4} - {\bar q}^{4}$ configurations are shown in Table~\ref{tab:table3}. 
The first line of ${\it q}^{5} - {\bar q}^{5}$ configuration in Table~\ref{tab:table3} shows the largest mass obtained within the ${\it q^3} - {\ \bar q^3}$
quark cluster assumption. Starting from the next line, we select the successive increasing masses, which fit well the experimental masses, again without any new adjustable parameter. They are, of course, many other masses predicted within this ${\it q}^{5} - {\bar q}^{5}$ quark configuration. The largest mass, corresponding to 5/2 for all four quantum numbers is M = 1331.7~MeV.

The agreement between data and predictions in Fig. 12 is noteworthy up to M = 250~MeV, and above 350~MeV. Three masses at M = 45.3, 81.3, and 231~MeV have no predicted counterparts. A mass, predicted at M = 323.7~MeV is not observed.                                                                                                                                                                                                                                                                                                                                                                                                                                                                                                                                                                                                                                                                                                                                                                                                                                                                                                                                                                                                                                                                                                                                                                                                                                                                                                                                                                                                                                                                                                                                                                                                                                                                                                                                                                                                                                                                                                                                                                                                                                                                                                                                                                                                                                                                                                                                                                                                                                                                                                                                                                                                                                                                                                                                                                                                                                                                                                                                                                                                                                                                                                                                                                                                                                                                                                                                                                                                                                                                                                                                                                                                                                                                                                                                                                                                                                                                                                               The pion mass and the structure at M = 252~MeV calculated at M = 249~MeV, are obtained at the same time with $q^{2}\bar q^2$ and ${\it q^3} - {\bar q^3}$   configurations.                                                                                                                                                                                                                                                                                                                                                                                                                                                                                                                                                                                                                                                                                                                                                                                                                                                                                                                                                                                                                                                                                                                                                                                                                                                                                                                                                                                                                                                                                                                                                                                                                                                                                                                                                                                                                                                                                                                                                                                                                                                                                                                                                                                                                                                                                                                                                                                                                                                                                                                                                                                                                                                                                                                                                                                                                                                                                                                                                                                                                                                                                                                                                                                                                                                                                                                                                                                                                                                                                                                                                                                                                                                                                                                                                                                                                                                                                                                                                                                                                                                                                                                                                                                                                                                                                                                                                                                                                                                                                           
The good agreement between data and calculations suggests to check if the same calculation with the same parameters $M_{0}$ and $M_{1}$ allows to reproduce larger masses observed in the $np \to np \pi^{+}\pi^{-}$ reaction \cite{troyan}.
Table 4 shows that this is indeed the case, using the  ${\it q}^{5} - {\bar q}^{5}$ quark configurations.

\begin{table}[h]
\begin{center}
\vspace*{-0.2cm}
\caption{Calculated masses within the ${\it q^5} - {\bar q^5}$ quark clusters, compared to the experimental values of
$M_{\pi^{+}\pi^{-}}$ \cite{troyan}.}
\label{tab:table4}
\vspace{0.mm}
\begin{tabular}{c c c c c c c}
\hline
conf.&$s_{1}$&$s_{2}$&$i_{1}$&$i_{2}$&M(MeV)&Exp.(MeV)\\
\hline
${\it q}^{5} - {\bar q}^{5}$&5/2&5/2&1/2&5/2&883.7&878$\pm$7\\
&1/2&1/2&5/2&5/2&1033&1036$\pm$13\\
&1/2&5/2&5/2&5/2&1182.3&1170$\pm$11\\
\hline
\end{tabular}
\end{center}
\end{table}

Such description is strengthen by the observation that the gap between two adjacent exotic meson masses are rather constant and equal to $\delta$M~$\approx$~18.7~MeV. A comparable situation was already observed in the studies of narrow exotic baryonic and dibaryonic masses. For example in the dibaryon field, between M = 2016~MeV, and 2194~MeV, five narrow masses were observed regularly separated by about 35.6~MeV. A mass gap of
$$\Delta M = 35~\mbox {~MeV~ } = 1/2 m_{e}/\alpha$$
where $m_{e}$ is the electron mass and $\alpha$ is the fine structure constant, was observed between several masses. Such gaps for leptons, mesons, and baryons were discussed since long time \cite{nambu,macgregor,palazzi}. 

A model was proposed  \cite{walcher},  which associates the narrow structure masses below the $\pi$ threshold production, to the multiproduction of "a genuine virtual Goldstone boson with a mass close to 20~MeV".  
This model can explain the level spacing of narrow mesonic structures experimentally observed.                                                                                                                                                                                                                                                                                                                                                                                                                                                                                                                                                             

It has been already mentioned that the usual calculations associate the multiquark clusters to masses 
larger than M = 1.5~GeV. The success of the present mass formula, based on quark clusters, which is able to reproduce many masses within a simple equation and only two parameters, suggests to consider the possibility of new physics on the basis of additional low mass hadrons. which would interact less strongly than the "classical" ones. 
\section{Conclusion}
The paper reviews  many data exhibiting narrow mesonic structures. Some are statistically well defined, not all of them.  These last data are considered when they appear in different publications, at close masses. The paper recalls also several data exhibiting narrow baryonic and dibaryonic structures. Detailed discussions concerning these last data were presented elsewhere, therefore they are only summarized in the present work.
It is shown that the masses of all these exotic structures have a common interpretation, namely that they correspond to a significant coupling of low mass mesonic narrow masses, with stable hadrons. 

The paper tentatively associates these masses to  stretched quark bags. A very simple phenomenological mass formula, allows to reproduce the experimental mesonic structure masses, and predict some masses of narrow structures at present not observed. These predictions concern narrow mesonic structures, as well as baryonic and dibaryonic structures.     

The effect of these narrow structures on $\pi\pi$ and NN phase shifts should be taken into account. Small variations from the present knowledge are anticipated. A qualitative discussion concerning $\pi\pi$ phase shifts was presented, together with similar remarks concerning NN phase shifts.



\begin{thebibliography}{99}
\bibitem{bormes}B. Tatischeff and E. Tomasi-Gustafsson, \emph{Search for Low Mass Exotic Mesonic Structures. Part I: Experimental Results.} Phys. Part. Nucl. Lett. {\bf 5}, (2008) 363;  Y. Yonnet {\it et al.}, \emph{ABC resonance in the ${\vec p}{\it p}$ $\to$ ppX$^{0}$ reaction, or is the ABC effect made of colored quark cluster configuration ?}, Phys. Rev. C{\bf 63}, 014001 (2001); B. Tatischeff {\it et al.}, \emph{Light mesons: $\pi$, $\eta$, and $\omega$ and exotic low mass meson production at intermediate energies in nucleon-nucleon scattering}, Phys. Rev. C{\bf 62}, 054001 (2000).
\bibitem{nelmum3}B. Tatischeff and E. Tomasi-Gustafsson, \emph{Exotic low mass unflavoured mesons: new data using old measurements}, to be published in Open Physics Journal.
\bibitem{troyan}Troyan Yu.A. {\it et al.}, \emph{Search and Study of Low-Mass Scalar Mesons in the Reaction np$\to$np$\pi^{+}\pi^{-}$ at the Impulse of Neutron Beam $P_{n}$ = (5.20 $\pm$0.12) GeV/c}, Physics of 
Part. and Nuclei Lett. 2012; 9: 77-87 (in russian).
\bibitem{pdg}J. Beringer ${\it et al.}$ (Particle Data Group), \emph{Review of particle Physics}, Phys. Rev. D {\bf 86}, 010001 (2012).
\bibitem{amsler}C. Amsler, \emph{Proton-antiproton annihilation and meson spectroscopy with the Crystal Barrel}, Rev of Modern Phys, 70: 1293-1339 (1998).
\bibitem{balestra}F. Balestra {\it et al.}, \emph{$\rho^{0}$ Meson Production in the ${\it pp} \to {\it pp\pi^{+}\pi^{-}}$ Reaction at 3.67~GeV/c}, Phys. Rev. Lett. {\bf 89}, 092001 (2002). 
\bibitem{aubert}B. Aubert {\it et al.}, (BABAR Collaboration), \emph{Precise Measurement of the $e^{+}e^{-} \to \pi^{+}\pi^{-}-(\gamma)$ Cross Section with the Initial State Radiation Method at {\it BABAR} [BABAR Collaboration]}, Phys. Rev. Lett. 103,  231801 (2009).
\bibitem{hehi}Alice: \emph{Physics Performance Report}, Volume II.  Phys. G.:Nucl Part Phys 2006; 32: 12951-747; Ippolitov M. and Vasil'ev A.  \emph{CALOR2004}, www.pg.infn.it/calor2004/program/.../ippolitov.pdf; Berger F. {\it et al.}, \emph{Particle identification in modular electromagnetic calorimeters}, Nucl. Instr. and Meth. in Phys. Research  A 1992; 321: 152-164.
\bibitem{abele}A. Abele {\it et al.}, \emph{${\bar p}{\it p}$-annihilation into $\omega \pi^{0}$,  $\omega \eta$ and $\omega \eta'$ at 600, 1200 and 1400~MeV/c},  Eur Phys J {\bf 12}, 429 (2000). 
\bibitem{prakhov}S. Prakhov {\it et al.}, \emph{Measurement of the invariant-mass spectrum for the two photons from the $\eta \to \pi^{0}\gamma\gamma$ decay}, Phys Rev C {\bf 78} 0152061 (2008).
\bibitem{nefkens}B.M.K. Nefkens {\it et al.}, \emph{New measurement of the rare decay $\eta \to \pi^{0}\gamma\gamma$ with the Crystal Ball/TAPS detectors at the Mainz Microtron}, Phys. Rev. C {\bf 90}, 025206 (2014), and arXiv:1405.4904v1 [hep-ex].
\bibitem{wilkin}A. Johansson and C. Wilkin, \emph{Hard bremsstrahlung in the $pp \to pp\gamma$ reaction.}, Physics Letters B{\bf 673}, 5 (2009); A. Johansson {\it et al.}, \emph{Measurement and analysis of the $pp \to pp\gamma$ reaction at 310~MeV}, arXiv:1101.5557v2 [nucl-ex] (2011).
\bibitem{abraamyan}Kh.U. Abraamyan {\it et al.}, \emph{Resonance structure in the $\gamma\gamma$ invariant mass spectrum in pC and dC interactions}, Phys. Rev. C{\bf 80}, 034001 (2009).
\bibitem{aachen}Aachen-Berlin-Bonn-Hamburg-Heidelberg-Munchen Collaboration (Desy), \emph{Photoproduction of Meson and Baryon Resonances at Energies up to 5.8~GeV}, Phys. Rev.  {\bf 175}, 1669 (1968).\\
\bibitem{zabrodin}A. Zabrodin {\it et al.}, \emph{Invariant mass distributions of nthe $\gamma n \to p \pi^{-} \pi^{0}$ reaction}, Phys. Rev.  C{\bf 60}, 055201 (1999).\\
\bibitem{langgartner} W. Langgartner {\it et al.}, \emph{Direct Observation of a $\rho$ Decay of the $D_{13}$(1520) Baryon Resonance}, Phys. Rev. Lett. {\bf 87}, 052001 (2001).\\
\bibitem{bilger}R. Bilger {\it et al.}, \emph{Two-pion production in proton-proton collisions near threshold at Celsius}, Acta Phys. Pol.{\bf B29}, 2987 (1998).
\bibitem{weis}M. Weis {\it et al.}, \emph{Separated cross sections in $\pi^{0}$ electroproduction at threshold at $Q^{2}$ = 0.05 GeV$^{2}/c^{2}$}, [A1] Collaboration], arXiv:0705.3816v1 [nucl-ex] (2007).
\bibitem{sirca}S. Sirca {\it et al.}, \emph{Structure of the Roper Resonance from Measurements of p($\vec{e}$,e'$\vec{p})\pi^{0}$ using Recoil Polarization Observables}, [E91011 Collaboration], http://www.ap.smu.ca/~sarty/RoperProp-Jan05.pdf.
\bibitem{frolov}V.V. Frolov {\it et al.}, \emph{Electroproduction of the $\Delta$(1232) Resonance at High Momentum Transfer}, Phys. Rev. Lett. {\bf 82}, 45 (1999).
\bibitem{patzold}J. P$\ddot{a}$tzold {\it et al.}, \emph{Study of the $pp \to pp \pi^{+} \pi^{-}$ Reaction in the Low-Enerfy tail of the Roper Resonance}, arXiv:nucl-ex/0301019v3 (2003).
\bibitem{bashkanov}M. Bashkanov {\it et al.}, \emph{Observation of a Structure in $pp \to pp \gamma \gamma$ near the $\pi \pi$ Threshold and its Possible Interpretation by $\gamma \gamma$ Radiation from Chiral Loops in the Mesonic $\sigma$ Channel},  arXiv:hep-ex/0406081v1 (2004); S. Kullander {\it al.}, \emph{First results from the CELSIUS/WASA facility}, Nucl. Phys. A {\bf 721}, 563c (2003).
\bibitem{maen}W. Maenner, \emph{Experimental Meson Spectroscopy}, 1974 (Boston), Proceedings of the fourth International Conference on Meson Spectroscopy, edited by D.A. Garelick, AIP Conf. Proc. No. 21  (A.I.P., New-York, 1974). 
\bibitem{ross}L. Rosselet {\it et al.}, \emph{Experimental study of 30 000 $K_{e4}$ decays}, Phys. Rev. D {\bf 15}, 574 (1977).
\bibitem{pisl}S. Pislak {\it et al.}, \emph{New Measurement of K$^{+}_{e4}$ Decay and the s-Wave $\pi\pi$-Scattering Length $a^{0}_{0}$}, Phys. Rev. Lett. {\bf 87},
221801 (2001).
\bibitem{shim}S. Shimizu {\it et al.}, \emph{Measurement of the $K^{+} \to \pi^{0}\pi^{0}e^{+}\nu(K^{00}_{e4})$ decay using stopped positive kaons}, Phys. Rev. D {\bf 70},
037101 (2004).
\bibitem{cola}G. Colangelo, \emph{Theory of $\pi\pi$ scattering}, arXiv:hep-ph/0011024v1 (2000). 
\bibitem{hoog}W. Hoogland {\it et al.}, \emph{Measurement and analysis of the $\pi^{+} \pi^{+}$ system produced at small momentum transfer in the reaction $\pi^{+}p \to \pi^{+} \pi^{+} n$ at 12.5~GeV},                Nucl. Phys. B{\bf 126}, 109 (1977).
\bibitem{kami}R. Kaminski, L. Lesniak, and B. Loiseau, \emph{Elimination of ambiguities in $\pi\pi$ phase shifts using crossing symmetry}, arXiv:hep-ph/0210334v1 (2002); 
\bibitem{kloe}W.M. Kloet and B. Loiseau, \emph{$\pi\pi$ scattering and the meson resonance spectrum}, Eur. Phys. J. A {\bf 1}, 337 (1998).
\bibitem{gray}G. Grayer {\it et al.}, \emph{High statistics study of the reaction $\pi^{-}p \to \pi^{-} \pi^{+}n$: Apparatus, method of analysis, and general features of results at 17~GeV/c},  Nucl. Phys. B{\bf 75}, 189 (1974).
\bibitem{hyam}B. Hyams {\it et al.}, \emph{$\pi\pi$ Phase-shift analysis from 600 to 1900 MeV}, Nucl. Phys. B{\bf 64}, 134 (1973).
\bibitem{pela}J.R. Pel$\acute{a}$ez and F.J. Yndur$\acute{a}$in, \emph{The pion-pion scattering amplitude}, arXiv:hep-ph/0411334v2 (2005).
\bibitem{borbar}B. Tatischeff {\it et al.}, \emph{Possible Evidence for Narrow States in Missing-Mass Spectra of the B = 2, T = 1 System}, Phys. Rev. Lett. {\bf 52}, 2022 (1984); B. Tatischeff {\it et al.}, \emph{Experimental evidence for narrow baryons in the mass range 1.0$\le$ M $\le$ 1.46~GeV}, Eur. Phys. J.A.{\bf 17}, 245 (2003); B. Tatischeff, \emph{Are the exotic mesons and baryons, recently observed, a signature of quark-hadron duality ?}, Ninth Int. Conf. on Hadron Spec. Protvino, AIP Conf. Proc. V 619, 765 (2001).
\bibitem{borbar1}B. Tatischeff {\it et al.}, \emph{$\pi$N and $\eta$p deexcitation channels of the N$^{*}$ and $\Delta$ baryonic resonances between 1470 and 1680~MeV}, Phys. Rev. c {\bf 72}, 034004 (2005).
\bibitem{borbar2}B. Tatischeff {\it et al.}, \emph{Possible indication of narrow baryonic resonances produced in the 1720-1790 MeV mass region}, Surveus in High Energy Physics, 19, 55 (2003).
\bibitem{btdou02}B. Tatischeff, \emph{Low mass exotic baryons: myth or reality ?}, Proc. of the XVI Int. Baldin Sem. on High Energy Phy. Problems, ISHEPP, V 2 153 (2002); B. Tatischeff and E. Tomasi-Gustafsson, \emph{Contribution to the study of narrow low mass hadronic structures}, arXiv:0802.0083v1 [nucl-ex] (2008).
\bibitem{filkov}L.V. Filkov {\it et al.}, \emph{New evidence for supernarrow dibaryons production in pd interactions}, Eur. Phys. J. A {\bf 12}, 369 (2001).
\bibitem{morsch}H.P. Morsch {\it et al.}, \emph{Radial Excitation of the $P_{11}$(1440~MeV) Resonance in Alpha-Proton Scattering}, Phys. Rev. Lett.{\bf 69}, 1336 (1992); Proc. of the Dixieme Journee Thematique de l'IPN d'Orsay (1995).
\bibitem{var06}B. Tatischeff and E. Tomasi-Gustafsson, \emph{Additional Evidence for Low Mass Exotic Baryons}, Proc. of the 11th Int. Conf. on Nucl. Rec. Mech., Varenna, 339 (2006).
\bibitem{laveissiere}G. Laveissiere {\it et al.}, \emph{Virtual Compton scattering in the resonance region up to the deep inelastic region at backward angles and momentum transfer squared of Q$^{2}$ = 1.0 GeV$^{2}$},  The Hall A Collaboration, arXiV:hep-ex/0406062 (2004).
\bibitem{var09}B. Tatischeff and E. Tomasi-Gustafsson, \emph{Charge exchange reactions on baryons and exotic low-mass narrow baryons}, Proc. of the 12th Int. Conf. on Nucl. Rec. Mech., Varenna, 455 (2009).
\bibitem{ellegarde}C. Ellegarde {\it et al.}, \emph{Reaction  (d$_{pol}, ^{2}He$) at Intermediate Energies}, Phys. Rev. Lett. {\bf 59}, 974 (1987); ibid \emph{Spin structure of the $\Delta$ excitation}, Phys. Lett. B {\bf 231}, 365 (1989); T. Sams {\it et al.}, \emph{Quasifree ($\vec{d},^{2}He)$ data}, Phys. Rev. C{\bf 51}, 1945 (1995); T. Sams. PhD thesis, Nils Bohr Institute Copenhagen (1991).
\bibitem{btjy}B. Tatischeff and J. Yonnet, \emph{Are narrow mesons, baryons, and dibaryons evidence for multiquark states ?}, Proc. Int. Work. "Relativistic Nucl. Phys. from hundreds MeV to TeV' Stara Lesna 194 (2000).
\bibitem{bordibar1}B. Tatischeff {\it et al.}, \emph{Evidence for narrow dibaryons at 2050, 2122, and 2150 MeV observed in inelastic {\it pp} scattering}, Phys. Rev. C{\bf 59}, 1878 (1999).
\bibitem{santi}L. Santi {\it et al.}, \emph{Evidence for narrow structure in the analyzing power of the $^{3}He(\vec{p},d)X$ reaction}, Phys. Rev. C {\bf 38}, 2466(R) (1988).
\bibitem{bordibar}B. Tatischeff {\it et al.}, \emph{Further Evidence Concerning narrow Isovector Nonstrange Dibaryons}, Europhys. Lett., {\bf 4} (6), 671 (1987).
\bibitem{khrykin}A.S. Khrykin {\it et al.}, \emph{Search bfor NN-decoupled dibaryons using the process $pp \to pp\gamma\gamma$ below the pion production threshold}, Phys. Rev. C {\bf 64}, 034002 (2001).
\bibitem{mul}P.J. Mulders, A.T. Aerts, and J.J. de Swart, \emph{ Multiquark states III ${\it Q^{6}}$ dibaryon resonances}, Phys Rev D {\bf 21} 2653 (1980); \emph{Multiquark states: ${\it Q^{3}}$ baryon resonances},  Phys Rev D {\bf 19} 2635 (1979); \emph{Negative-Parity NN Resonances and Extraneous States}, Phys. Rev. Lett. {\bf 40}, 1543 (1978).
\bibitem{nambu}Y. Nambu, \emph{An empirical mass spectrum of elementary particles}, Prog. Theor. Phys. {\bf 7}, 595 (1952).
\bibitem{macgregor}M.H. Mac Gregor, \emph{Electron generation of leptons and hadrons with reciprocal alpha-quantized lifetimes and masses}, Int. J. Mod. Phys. {\bf A20} 719 and 2893 (2005).
\bibitem{palazzi}P. Palazzi, \emph{Are Hadrons Shell-Structured ?}, Proc. $8^{th}$ Int. Symp. "Frontiers of Fundamental Physics", Madrid (2006); Proc. $9^{th}$ Int. Work. on Meson production, Krakow, (2006).
\bibitem{walcher}T. Walcher, \emph{A simple model to explain narrow nucleon resonances below the $\pi$ threshold}, arxiv:0111279 [hep-ph].


\end{thebibliography}
\end{document}